\documentclass[twocolumn,pra,aps,longbibliography]{revtex4-1}
\usepackage{amsmath}
\usepackage{graphicx}
\usepackage{sistyle}
\usepackage{color}
\usepackage{float}
\usepackage{enumitem}
\usepackage{todo}

\begin{document}

\title{Perfectly secure steganography: hiding information in the quantum noise of a photograph}
\author{Bruno Sanguinetti}
\email{Bruno.Sanguinetti@unige.ch}
\address{Group of Applied Physics, University of Geneva, Switzerland}
\author{Anthony Martin}
\address{Group of Applied Physics, University of Geneva, Switzerland}
\author{Giulia Traverso}
\address{Fachbereich Informatik, Technische Universit\"at Darmstadt, Germany}
\author{Jonathan Lavoie}
\address{Group of Applied Physics, University of Geneva, Switzerland}
\author{Hugo Zbinden}
\address{Group of Applied Physics, University of Geneva, Switzerland}

\begin{abstract}
We show that the quantum nature of light can be used to hide a secret message within a photograph. Using this physical principle we achieve information-theoretic secure steganography, which had remained elusive until now. The protocol is such that the digital picture in which the secret message is embedded is perfectly undistinguishable from an ordinary photograph. This implies that, on a fundamental level, it is impossible to discriminate a private communication from an exchange of photographs.
\end{abstract}
\maketitle

\section{Introduction}
Communication is crucial in human society. As we become more connected, our ability to safeguard and protect our communications becomes critical. Although privacy of correspondence is a right granted by the constitution of most countries~\cite{King2012}, technical solutions must guarantee this right in practice. Cryptography~\cite{Bauer2000} aims at making a message incomprehensible to the unauthorized reader, however it does not guarantee privacy: the fact that an encrypted message is being exchanged can be discovered by observing encrypted (random) data within a communication.

Steganography provides a solution by concealing the existence of the message in an innocent support. Historically, Histaeus, tyrant of Miletus, instructed Aristagoras to revolt by sending him a message tattooed on a slave's scalp, under his hair. Later, stegosystems continued to play a historical role, notable examples are the invisible inks used during the American Revolution and the microdots used in World~War~II. Nowadays, steganography often looks at hiding information in a digital image~\cite{Johnson1998,Poornima2013}.

Today, governments are proposing compromises between privacy and security, exploring the idea of prohibiting unbreakable encryption protocols. In this context, the question of whether such a protocol can be fundamentally undetectable is important: if a perfectly secure steganographic scheme exists, any attempt at prohibiting secret communications, e.g. by weakening cryptographic protocols, is vain. We consider the following setting: Ward is a security agent examining an innocent-looking photograph sent by Alice to Bob. Ward has to decide whether this photograph contains a secret message or not. Is it, at least in principle, possible for Ward to make this decision?

Although it has been shown that perfectly secure steganography is in principle possible, a protocol has not been proposed~\cite{Anderson1998,Hopper2002}.
Here we show that the quantum nature of light~\cite{Planck1900,Einstein1905} can be used to perfectly hide the existence of a secret message in a photograph.

First, we will give the context and setting for steganography, followed by an overview of the state of the art and its limits. We then describe a protocol that uses the presence of shot noise (quantum noise) within a photograph to hide information perfectly.
A proof-of-principle experiment is shown in the Supplementary Material.

\section{Digital image steganography}

\begin{figure}[htbp]
\begin{center}
\includegraphics[width=\columnwidth]{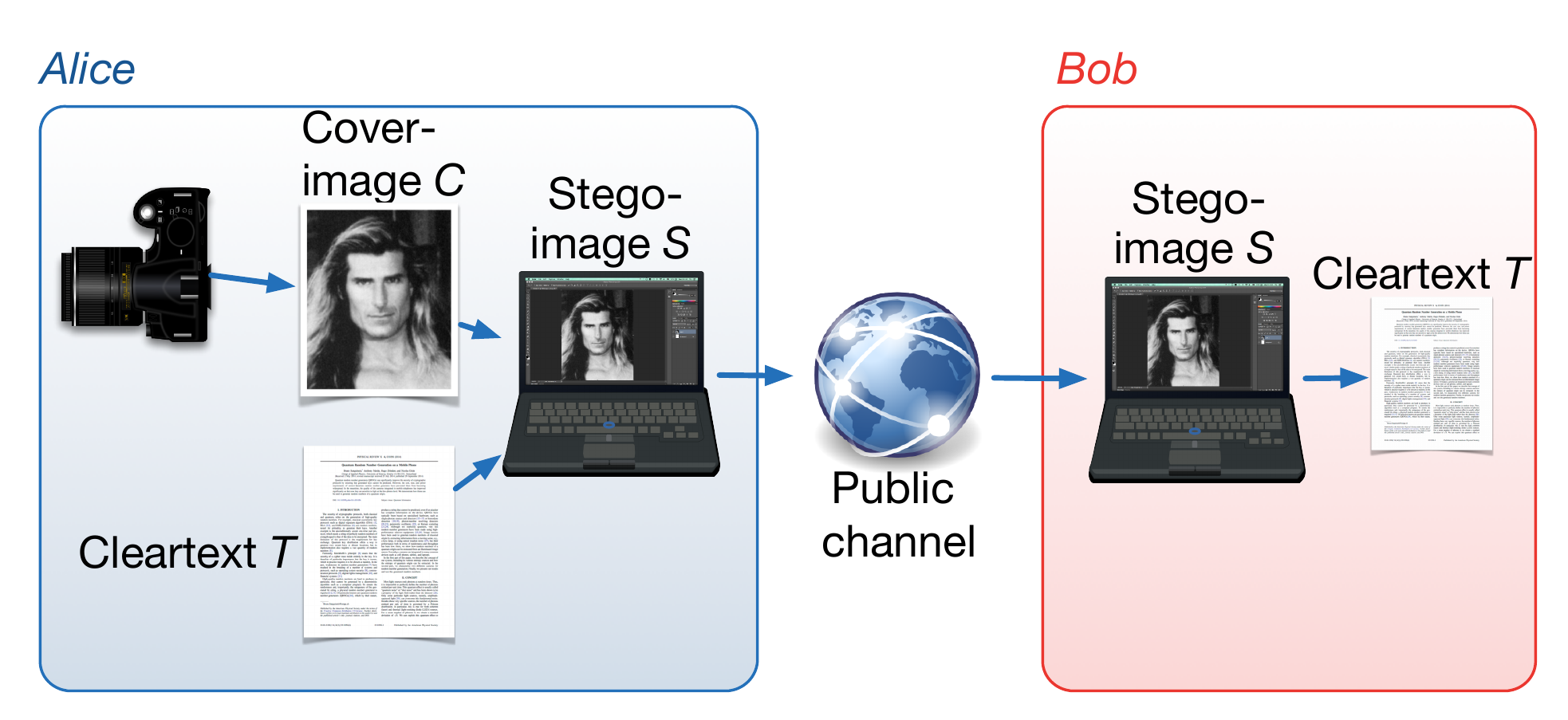}
\caption{Illustration of a generic image steganography protocol: Alice embeds the cleartext in a photograph to form a stego-image. This image is published where Bob can find it. Bob then extract the cleartext from the image.}
\label{fig:Steganography_concept}
\end{center}
\end{figure}

As described in Fig.\,\ref{fig:Steganography_concept}, steganography on digital images is nowadays mostly based on embedding a secret message $T$ into a \emph{cover-image} $C$. A cover image is an innocent-looking digital image used to hide the secret content to protect. When this image $C$ contains the secret message $T$, i.e. once the embedding of $T$ has been performed, then it is referred to as \emph{stego-image} and denoted by $S$. The image $S$ can then be distributed over the Internet without arising suspicion ~\cite{Poornima2013,Anderson1998,Cheddad2010}. The secret message is embedded using the \emph{embedding key}. Once the receiver gets the stego-image $S$, it can retrieve the hidden message $T$ using the corresponding \emph{extraction key}. The case where the embedding key is the same as the extraction key is referred to as \emph{secret-key steganography}. Instead, in \emph{public-key steganography} the embedding key and the extraction key are not identical. Aiming at information-theoretically secure steganography, in our work we consider secret-key steganographic schemes only. In fact public-key steganography is information-theoretically impossible, as proven in~\cite{Ahn2004}. \\

In our setting, Ward has to decide whether a certain digital image contains an embedded message or not. This means that Ward is dealing with a hypothesis test problem, as pointed out by Cachin in~\cite{Cachin1998}. A steganographic scheme is perfectly secure if Ward cannot detect the presence of any message which has been embedded with this scheme.

This can only happen if the stego-image $S$ is drawn from the same statistical distribution as the cover-image $C$. If that is the case, then it is referred to as a perfectly secure steganographic scheme.  However, the embedding of any object will inevitably modify the statistical distribution of the hosting digital image. 
 
Although the distortions due to embedding cannot be perceived by human eyes, they can be detected from a statistical point of view. More precisely, let us denote with $\cal{C}$ the statistical distribution of the pixels in the cover-image $C$ and with $\cal{S}$ the one for the stego-image $S$. As the picture $S$ has been obtained through embedding from the picture $C$, then the statistical distribution $\cal{S}$ diverges from $\cal{C}$. The distance between these two distributions $\cal{C}$ and $\cal{S}$ is quantified by the Kullback-Leiber divergence~\cite{Cachin1998}, denoted by $D_{KL}(\cal{C}\vert \vert \cal{S})$. In the discrete case, which is our case of interest, this value is defined as:

\begin{center}
        $D_{KL}(\cal{C}\vert \vert \cal{S}):=$ $ \sum_{i} {\cal{C}}(i)$ $ \log{\frac{{\cal{C}}({i})}{{\cal{S}}({i})}} $.
\end{center}
        
\begin{figure}[htbp]
        \begin{center}
                \includegraphics[width=\columnwidth]{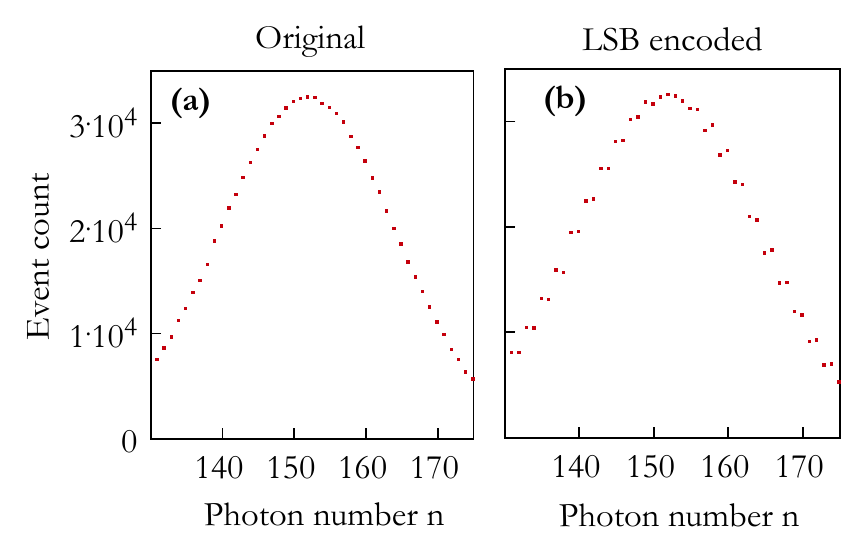}
                \caption{Histogram of the pixel values of a homogenous area of a photograph (a) and the obvious effects of encoding encrypted data on the Least-Significant Bit of each pixel (b). The effect is visible even if the probability of the least significant digit being 1 is exactly 0.5.}
                \label{fig:lsb_gaussian}
        \end{center}
\end{figure}

Specifically, the Kullback-Leiber divergence of $\cal{S}$ from $\cal{C}$ is the relative entropy between the two distributions, as it measures the loss of information when $\cal{S}$ approximates $\cal{C}$. Performing the embedding leads to a Kullback-Leiber divergence strictly greater than zero. An example is shown in Fig. \ref{fig:lsb_gaussian}, where the distortion introduced by the technique of the Least-Significant-Bit (LSB) replacement is immediately visible. For an accurate description of this technique we refer to ~\cite{Cheddad2010}. Other common steganographic techniques can be found in Refs.~\cite{Barni2001} and ~\cite{Boato2010}.

In some approaches (see ~\cite{Solanki2006,Sullivan2006}), zero Kullback-Leiber divergence is achieved at the expense of employing an always larger amount of hosting signal (i.e. in the cover-image $C$) for statistical restoration. However, these approaches do not scale: for a constant risk of detection, the amount of data that can be hidden is proportional to the square-root of the total size of the image material sent~\cite{Ker2008}, so that if Alice sends a constant flow of images to Bob, the amount of data that she can embed per image quickly goes to zero.
 
Instead of embedding, i.e. attempting to emulate the original statistical distribution of the cover image, which necessarily leads to distortion, our scheme creates the stego-image $S$ by directly sampling the space $\cal{C}$ of cover images, ensuring zero distortion.

\section{Protocol} 
We propose a secret-key steganographic protocol where two digital pictures are involved. This is in contrast with the common steganographic strategies where the protocol takes into account one digital picture only, i.e. the cover-image. In our framework, one picture is the \emph{key-image} $K$, and the other is the cover-image $C$, which is discarded as soon as the stego-image $S$ is created. The randomness in our protocol is given by the shot noise (quantum noise) naturally present in all images~\cite{Sanguinetti2014}.

We take the point-of-view of Ward, and assume that Alice has the capability of satisfying the following assumptions:
\begin{itemize}
        \item[\emph{i})] The state of the camera and the subject remain unchanged between the taking of two consecutive photographs.
        \item[\emph{ii})] Each pixel is statistically independent, i.e. one can not predict the value of one pixel by the knowledge of the others better than the shot noise limit.
\end{itemize}

\begin{figure*}[htbp]
\begin{center}
\includegraphics[width=15cm]{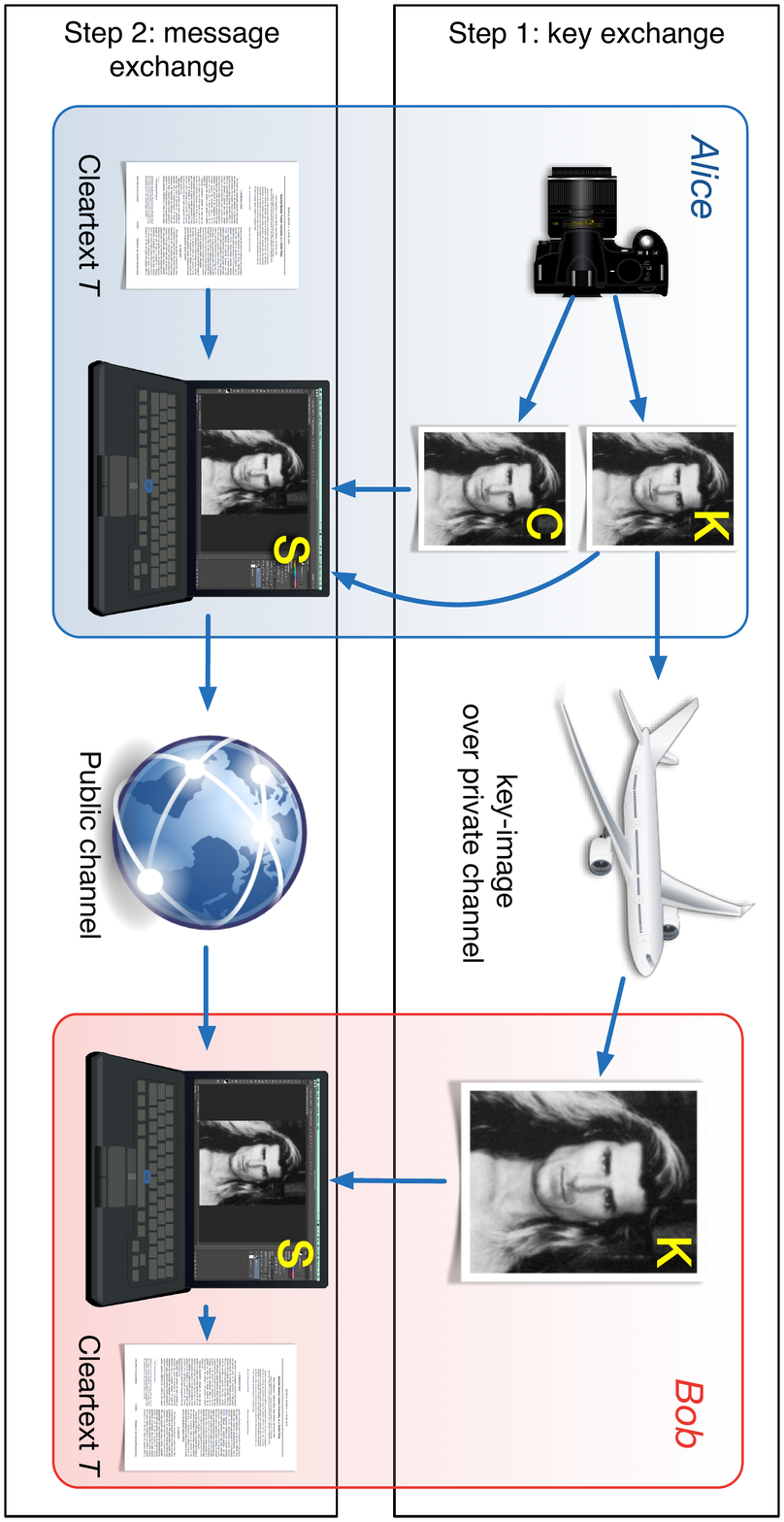}
\caption{Illustration of the proposed protocol: first, Alice and Bob meet. Alice takes two ``identical'' photographs $K$ and $C$ and gives a copy of $K$ to Bob as a key. They then separate. At a later date, Alice encodes her cleartext with the help of the key-image $K$ and the cover-image $C$ into a stego-image $S$. She then published this stego-image where Bob can retrieve it. With $K$ and $S$ in hand, Bob is able to extract the cleartext from $S$. }
\label{fig:Q_steganography_concept_g}
\end{center}
\end{figure*}

In the following we describe the protocol, also illustrated in Fig.~\ref{fig:Q_steganography_concept_g}. Alice wants to secretly communicate a message $T$ to Bob. Alice takes two photographs, $K$ and $C$. In order to fulfill assumption (\emph{i}), she uses the same camera and takes the pictures of the same static subject in rapid succession. As this is a secret-key steganographic protocol, the photograph $K$ is shared with Bob over a private channel (for example, they can meet in person). According to Kerckhoffs's principle~\cite{Kerckhoffs1883}, the protocol is aborted if the key-image $K$ is seen by Ward. \\

The $i$-th bit of the message $T$ is denoted by $T_i$, while the $i$-th pixel value of the images $K$, $C$, and $S$ is denoted by $K_i$, $C_i$, and $S_i \in \mathbb{N}_0$, respectively. The secret message $T$ is encoded within the stego-image $S$. The pixels composing such image $S$ are taken either from $K$ or $C$, according to the following rule:
\begin{equation}
        S_i := \begin{cases}
        K_i, \quad \text{if} \quad T_i=0\\
        C_i, \quad \text{if} \quad T_i=1
        \end{cases}
\end{equation}

Once the stego-image $S$ is constructed, Alice sends it to Bob through a public channel. The cover-image $C$ is then destroyed. On his side, Bob can decode the hidden message $T$ as he has the secret-key $K$. He retrieves $T$ bit by bit checking the pixels $S_i$ of the stego-image $S$ he received by applying:
\begin{equation}
T_i := \begin{cases}
0, \quad \text{if} \quad S_i = K_i\\
1, \quad \text{if} \quad S_i \neq K_i.
\end{cases}
\end{equation}

In practice, there exists the possibility that $K_i=C_i$, which leads to a bit error if $T_i=1$. To solve this issue, $T$ can already contain the required error correction information, such that it will not interfere with the rest of the procedure. For a more detailed explanation we refer to the Supplementary material and the proof-of-principle experiment.

\section{Security Analysis}

The stego-image $S$ is a digital picture that \emph{could} have been drawn directly from the underlying distribution of the genuine pictures $K$ and $C$. The photographs $K$ and $C$ naturally follow the same statistical distribution that we denote by $\cal{C}$. That is, the statistical distribution of the key-image and the cover-image is seen as the statistical distribution of the cover-image of the usual steganographic framework~\cite{Cheddad2010}. The statistical distribution of the stego-image $S$, denoted by $\cal{S}$ do not deviate from $\cal{C}$. That is possible because of Assumptions (\emph{i}) and (\emph{ii}) stated at the beginning of the protocol. This means that for each sample $i$ of the statistical distributions, it holds that
$
{\cal{C}}(i) = {\cal{S}}(i).
$
This leads to a zero Kullback-Leiber divergence, meaning that the message $T$ is carried by the stego-image $S$ without statistical distortions with respect to $C$, i.e. the secret content is perfectly undetectable by Ward. This leads to a perfectly secret steganographic scheme. Furthermore, all the pixels available are used to encode the message $T$, which has the same size of the the images $K$, $C$, and $S$. Contrary to the common steganographic strategies, our protocol the amount of the data embedded scales linearly with the image size.

Moreover, the protocol that we propose bares strong similarities to the one-time pad (OTP). In OTP, to encode a bit '0', Alice choosing a bit from the key $k$, to encode a bit '1' Alice chooses a bit from $\bar{k}$. We replace $k$ and $\bar{k}$ by the two images $K$ and $C$. Our protocol therefore has the advantages, but also the requirements of the OTP. In the next section, we show this formally, and generalize it to any provably-secure steganographic protocol.

\section{Perfect steganography implies perfect cryptography}

The definition of perfectly secure stegosystems proposed by Cachin~\cite{Cachin1998} parallels the one of perfectly secret cryptosystems proposed by Shannon~\cite{Shannon1949}. 
Below, we show that any perfectly secure steganographic protocol is also a perfectly secret cryptographic protocol, and must therefore have the same (or greater) requirements in terms of key length and randomness.

Indeed, if it is impossible to detect the presence of a message in the cover-image, it is also impossible to extract the message without knowledge of the key. This can be proven by \emph{reductio ad absurdum}: if Ward has access to an eavesdropping function, he can use it to obtain the message and therefore discover its presence. Perfect steganography implies that it is impossible to detect a hidden message, and consequently this implies that an eavesdropping function can not exist. 

More formally: Let us suppose that $\cal{T}$ is the space of the messages to hide and $\cal{S}$ the space of the stego-images. A steganographic protocol is a function $P: \cal{T} \to \cal{S}$ such that $P(T)=S$, for a certain $S \in \cal{S}$ and a certain $T \in \cal{T}$. The space of the steganographic protocols is denoted by $\cal{P}$. In this framework, a hypothesis test is a function $H: \cal{S} \to $ $\{0,1\}$ defined as follows:
$$
H(S):= \begin{cases}
1 \quad \text{if there is a message within } S \\
0 \quad \text{otherwise}.
\end{cases}
$$
Let us suppose that a message $T \in \cal{T}$ has been embedded using the protocol $P \in \cal{P}$, generating the stego-image $S \in \cal{S}$. Furthermore, let us assume that the eavesdropper Ward is able to extract the message $T$. This means that there exists an eavesdropping function $E_{P} \in \cal{E}$ able to break the protocol $P$. More precisely,
\begin{center}
        $\exists  E_{P} \in \cal{E} \quad \text{such that}  \quad $ $ E_{P}(S)= T, $
\end{center}
where ${E}_{P}: \cal{S} \to \cal{T}$, from which Ward can infer that $H(S)=1$. However, a perfectly secret steganographic scheme $R \in \cal{P}$ is perfectly undetectable, so that for any stego-image $S \in \cal{S}$ generated using $R$ behaves like a common image, so that for a $T \in \cal{T}$
$$
S=R(T) \Rightarrow H(S)=0.
$$
This is in contrast to the result $H(S)=1$, which Ward can infer if $E_{P}$ exists, implying that $E_{P}$ cannot exist, and any perfectly secure steganographic scheme is also a perfectly secure cryptographic scheme.

\section{Experiment}

An image sensor (CCD or CMOS) converts photons that impinge on its pixels into electrons. Each absorbed photon will generate a single electron. This charge is converted into a voltage and digitised, so if there were no technical noise, the digital values are a direct representation of the number of photons absorbed by the pixel. Photons are emitted by the light source illuminating the observed object at an unpredictable time, due to the laws of quantum physics. The number of photons absorbed by a sensor pixel during the exposure, follows the Poisson distribution, and is fundamentally random. Therefore, the standard deviation of the measured number of electrons (photons) $n$ will be $\sqrt{n}$. For most imagers, the full-well capacity can be of $\SI{5e4}{}$, so $n\sim\SI{e4}{}$, and the standard deviation will typically be $\sigma\sim100$.
This is much larger than the noise levels of modern image sensors, which is most often $<10\,e^-$, and can be of the order of a single electron in devices with small pixels, such as the cameras of mobile telephones.
The noise in most photographs is therefore dominated by quantum noise~\cite{Sanguinetti2014}. This type of noise arises from the quantum nature of light and is omnipresent, the only exception being ``squeezed light''~\cite{Walls1983}, which can only be created in complex quantum-optical experiments. If the images were noise-free, the protocol that we propose would not work, as the key-image and cover-image would be identical.

We have performed the above protocol in two ways. On one hand, we used a scientific monochrome camera. Although unrealistic from a practical perspective, this allows us to explore the theoretical framework without color image processing. On the other hand, we used a consumer color camera, which produces raw image files. Alice's pixel manipulations are carried out at the raw image stage, although the image can be later processed.

These experiments, which are detailed in the Supplementary Material, show that indeed quantum noise is the dominant noise mechanism for both cameras. Furthermore, we found that adjacent pixels are statistically independent, showing that Assumption (\emph{ii}) can be satisfied~\cite{Sanguinetti2014}. For the scientific monochrome camera, the image-to-image fluctuations were smaller than both the shot noise and the typical efficiency fluctuations between pixels, indicating that Assumption (\emph{i}) can be satisfied. With the commercial color camera, fluctuations in the shutter speed introduced a measurable difference between consecutive images. A threshold should be derived to evaluate what fluctuations are acceptable for our assumptions to hold. Further details are given in the supplementary material, in particular, using more than one pixel to encode each bit strongly increases the robustness of the protocol with respect to image manipulation and experimental imperfections. Intuitively, the protocol is perfectly robust in the limit where Alice and Bob use an entire image to encode each bit. 

Our experimental demonstration illustrates how the proposed protocol can be carried out. We show that Ward cannot in principle know whether an image contains hidden information, however, from Alice and Bob's perspective, further study is required to find all possible problems and loopholes in the physical implementation which could be exploited by Ward.

\section{Conclusion}
We have shown that the answer to the question ``can Ward, at least in principle, make the decision of whether a photograph contains any hidden information?'' is ``No''. More precisely, we demonstrate that a provably secure stegosystem is possible and propose a concrete protocol. This proves that a photograph can carry a large amount of hidden data whilst being indistinguishable, in a fundamental way, from a typical photograph. It also shows that an unmodified photograph can be used as the key. We show that any perfectly secure stegosystem is also a perfectly secret cryptosystem, and has at least the same requirements, i.e. the key must be at least as long as the message and must be perfectly random. Randomness is  taken directly from the shot noise (quantum noise) which is dominant in modern digital cameras. We conclude that it is impossible to prohibit encrypted communications without prohibiting the free exchange of photographs.

\section{Acknowledgements}
We acknowledge the Swiss NCCR QSIT for financial support. This work has been co-funded by the European Union's Horizon 2020 research and innovation program under Grant Agreement No 644962.  J. L. was supported by the Natural Sciences and Engineering Research Council of Canada (NSERC).
We thank Deanna Needell, Fabio, and Evgeniya Balysheva for providing alternative photographs to the commonly used picture of ``Lena''. We are also  grateful to Nicolas Gisin for useful discussion.

\bibliography{steganography_bibliography.bib}
\bibliographystyle{apsrev}

%
%

\appendix

\section{Proof-of-Principle experiment}
We have performed the above protocol in two ways. On one hand, we used a scientific monochrome camera. Although unrealistic from a practical perspective, this allows us to explore the theoretical framework without color image processing. On the other hand, we used a consumer color camera, which produces raw image files. Alice's pixel manipulations are carried out at the raw image stage. Note that strong error correction is needed in case the stego-image is compressed into a JPEG. If that is the case, also the algorithm on Bob's side becomes more complex.

\subsection{Experimental Setup}
The experimental setup consists of a monochrome scientific camera (ATIK 383L+)  mounted to an optical bench. The noise of this camera is strongly dominated by quantum shot noise, as it has been discussed in~\cite{Sanguinetti2014}. Moreover, we show that the statistical distribution of pixel values are independent, meaning that Assumption (\emph{ii}) is satisfied. With respect to the pictures $K$ and $C$, the subject has to be static while they are taken. Any variation of the experimental conditions would conflict with a (\emph{i}). The subject is a printed circuit board (PCB) with some areas of strong contrast between the reflective copper traces and dark board color. This saturates some pixels and gives a predictable value, leading to errors occurrence. These errors are useful for testing our error correction algorithm.
To satisfy Assumption (\emph{i}), we used a DC-powered light-emitting diode (LED), which provides a constant intensity level. 

Exposure time of 100 ms  ensures that each pixel, even the dark ones, receive a sufficiently large photon number, such that the quantum noise dominates other possible noise types.
A set of 100 photographs is taken in order to verify the stability of the setup and the repeatability of the measurements. All the photographs are 8 megapixels, encoded as 16 bit ``tiff'' files. Among these photographs, only two of them are finally chosen and actually used to perform the protocol. Specifically, the first one is employed as the key-image $K$, while the second one as the cover-image $C$.

\subsection{Protocol Implementation in Monochrome}
\begin{figure*}[htbp]
\thispagestyle{empty}
\centering
\includegraphics[width=2\columnwidth]{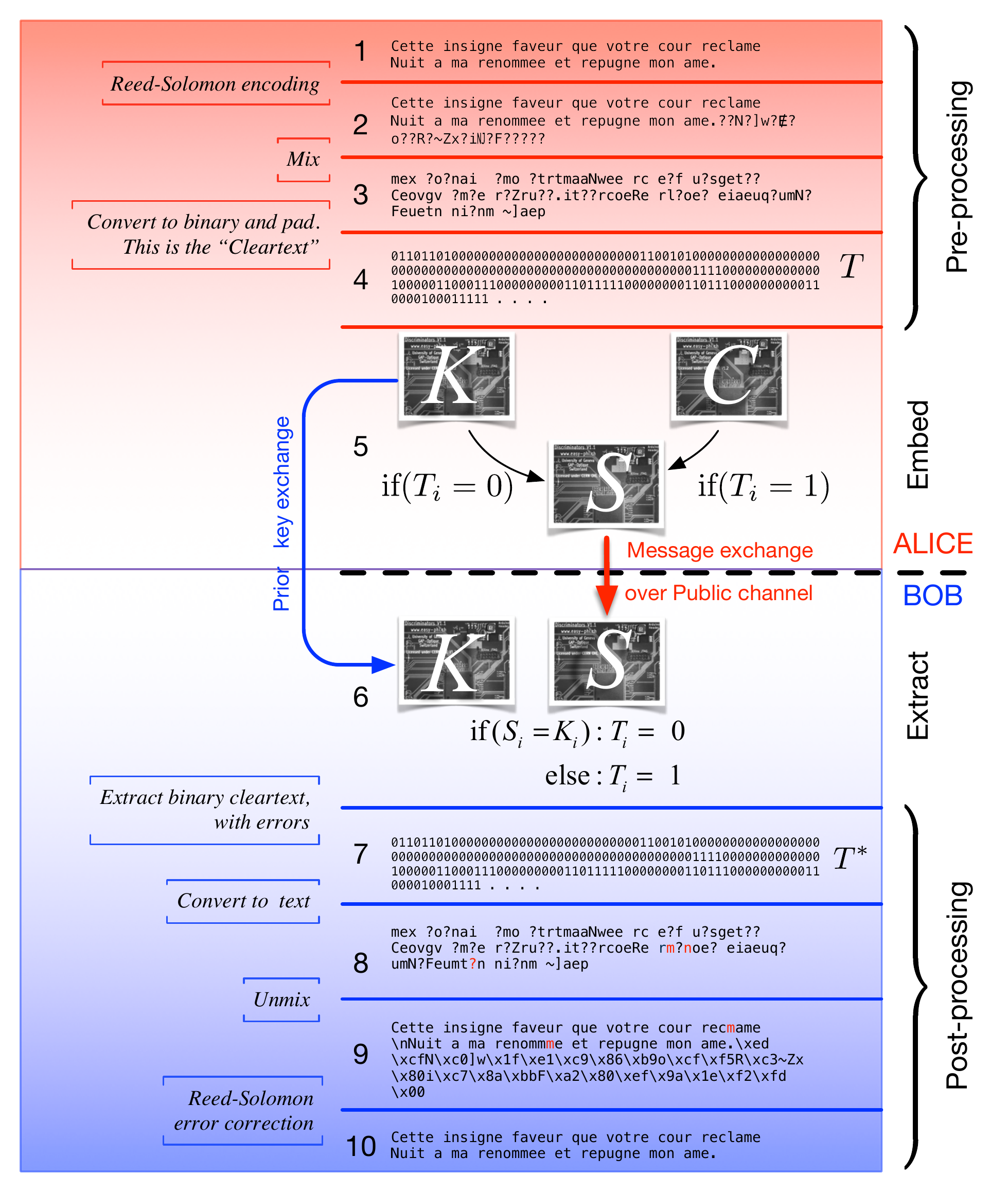}
\caption{Illustration of the full algorithm. The original text (1) is Reed-Solomon encoded (2). Each character is placed at a pseudorandom position in the image with a mixing step (3). The data is converted to binary and padded as to fit the image extents (4). All these steps are ``pre-processing''; we consider the binary string obtained after step (4) as being the cleartext. Step (5): steganographic encoding by choosing each pixel of the stego-image $S$ from either $K$ or $C$ depending on the cleartext value being 0 or 1 respectively.
The following steps are performed by Bob. He compares $S$ with his key-image $K$, to recover the cleartext $T^*$ (6). Note that step (5) cannot be fully reversed by step (6), due to the probability that $K_i=C_i$. The extracted $T^*$ therefore contains errors, which are recovered during  a Reed-Solomon step (9) to yield the original text (10).
}
\label{fig:programme_illustration}
\clearpage
\end{figure*}

The following protocol was implemented using the Python programing language. A representation of the functions, with real sample data, is shown in Fig.~\ref{fig:programme_illustration}.

\begin{figure*}[htbp]
\begin{center}
\includegraphics[width=15cm]{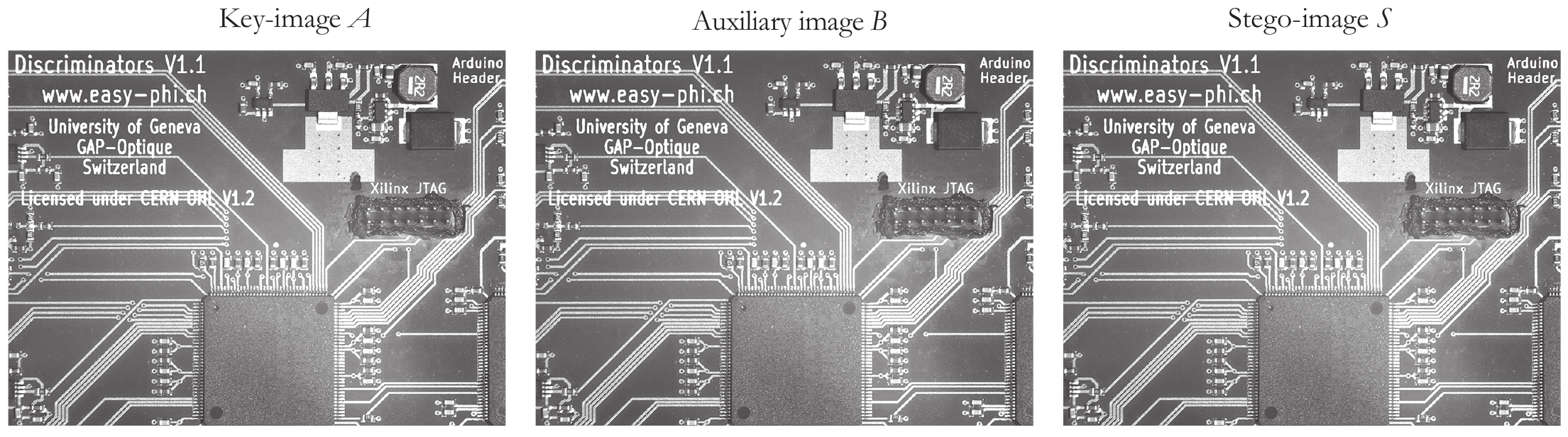}
\caption{The three actual photographs as used in the proof-of-principle experiment. A printed-circuit board was chosen as it goes towards satisfying our assumption of a static subject and presents a high-contrast image with some saturated pixels, which will therefore have the same value in pictures $K$ and $A$, and test the error-correction code.}
\label{fig:Three_PCB_photos}
\end{center}
\end{figure*}

The protocol consists of the ten steps listed below. It is assumed that Alice has already taken two photographs $K$ and $C$ and securely shared $K$ with Bob.
\begin{enumerate}
\item Alice choses the text to communicate to Bob. The size of the text is at most one bit per pixel. In our case this would be a maximum of \SI{8}{Mbits}.
\item Alice encodes the text using the Reed-Solomon code~\cite{Wicker1999}. The required redundancy depends on the image, and more or less space should be allocated for error correction. In our tests, the error rate is approximately 1\%, due mostly to overexposed pixels. This requires the allocation of at least 2\% of the total space for error correction.
Note that Alice can herself test whether the redundancy is sufficient or not, because she can locally perform all Bob's operations.
\item Alice places each byte of data in a specific region of the image. The place where each byte is allocated is predetermined but looks pseudorandom within the image. The placement list is obtained by shuffling an ordered dictionary of the placement of each byte. The shuffled dictionary is previously shared with Bob and can be a constant. The purpose of this ``mixing'' step is to optimise error correction efficiency. Often, overexposed pixels are adjacent and this would damage many bytes within a Reed-Solomon block, making the data unreadable. The mixing step allows for a homogeneous distribution of the bytes within a block. This also leads to a homogeneous distribution throughout the image itself, mixing even a larger overexposed area recoverable by means of error correction.
\item The data is converted to its binary representation and padded such that there is exactly one bit per pixel. The output of this step corresponds to the message $T$ introduced in the paper.
\item The stego-image $S$ is created by $K$'s pixels where $T=0$ and $C$'s pixels where $T=1$. Alice communicates the picture $S$ to Bob through a public channel, without raising suspicions.
\item Bob is already in possession of the key-image $K$ and receives the stego-image $S$. He retrieves the message $T^*$ by comparing each pixel pair $(S_i, K_i)$. He sets $T^*_i=0$ if $S_i = K_i$ and $T^*_i= 1$ otherwise.
\item In the extracted message $T^*$ errors occur in the positions where $K_i=A_i$. Error correction is then needed.
\item Bob records the bytes according to his dictionary. Note that, unlike the error-locations, the dictionary can be public. The dictionary does not have to be carried by Bob, and will therefore not reveal his intention to communicate to Ward.
\item Bob performs the Reed-Solomon error correction (errors are highlighted in Red in Fig.~\ref{fig:programme_illustration}.)
\item Bob gets the original text without errors.
\end{enumerate}
We have performed the above protocol using two images of a printed circuit board. The results are shown in Fig.~\ref{fig:Three_PCB_photos}.

\begin{figure}[htbp]
\begin{center}
\includegraphics[width=\columnwidth]{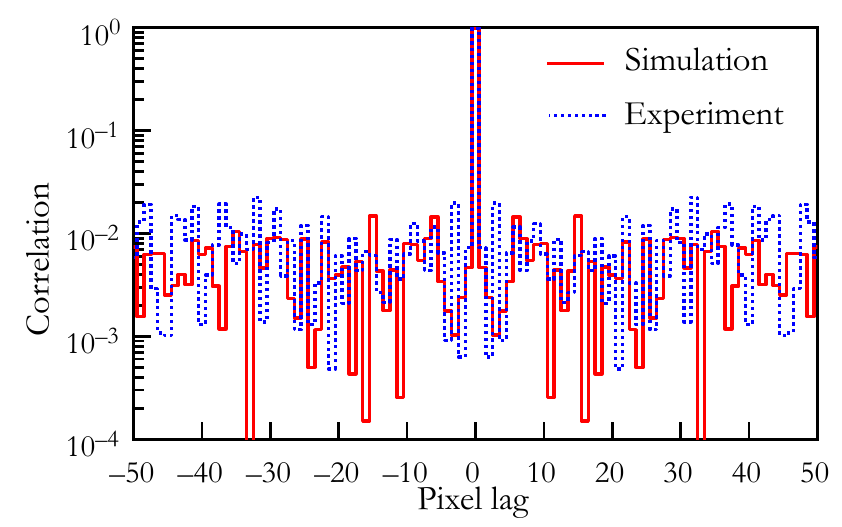}
\caption{Autocorrelation between adjacent pixels. ``Pixel lag'' is the distance between pixels. Here, the autocorrelation is calculated over the subtraction of two consecutive images, to cancel out any static image feature. The curve labeled ``theory'' represents the same process applied to images generated using a pseudorandom Poisson distribution, and are used to evaluate the finite-sample effects. }
\label{fig:autocorrelation}
\end{center}
\end{figure}

\begin{figure*}[htbp]
\begin{center}
\includegraphics[width=12cm]{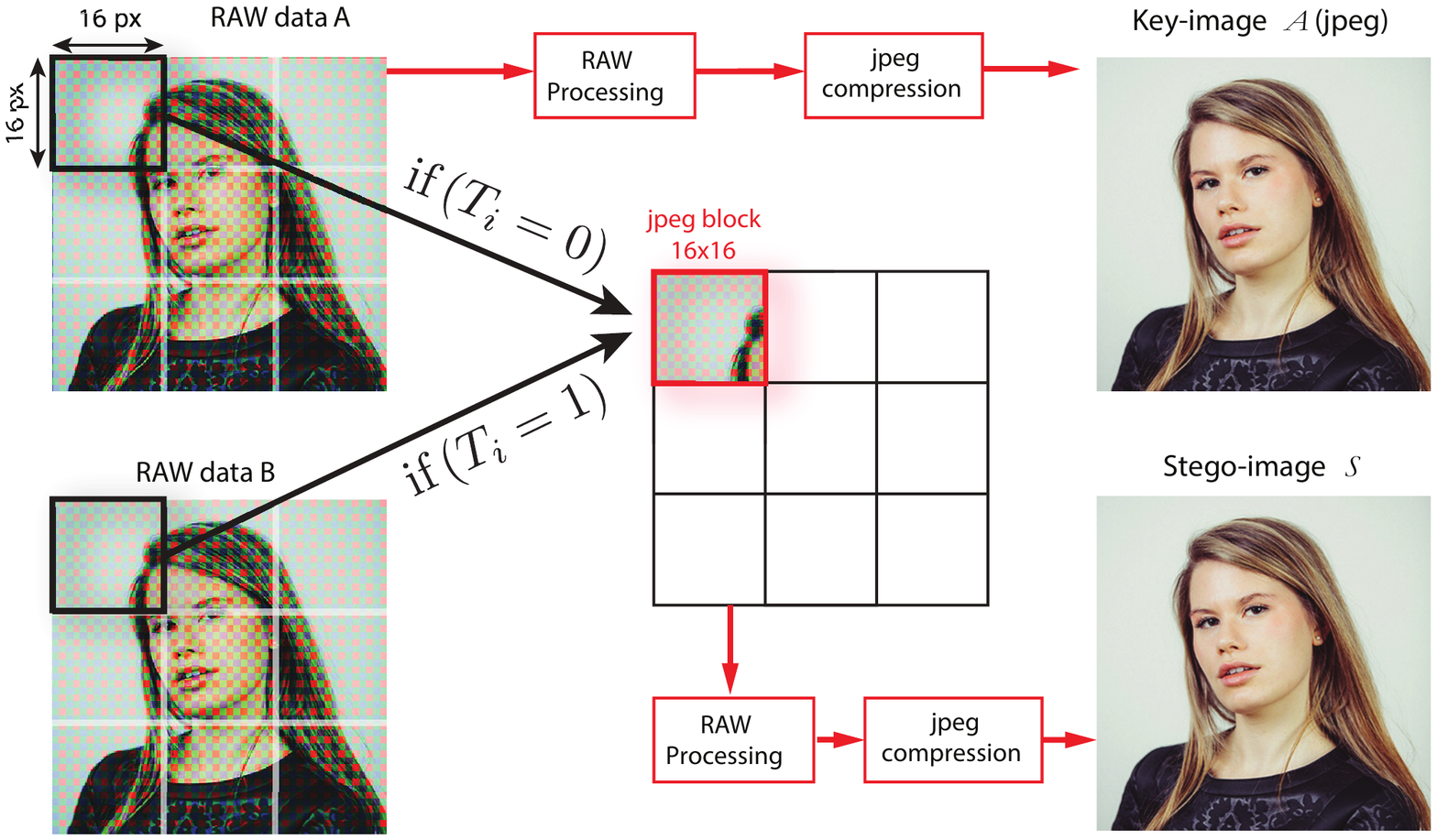}
\caption{Illustration of how the protocol might work for jpeg-compressed images: instead of using the images pixel by pixel, the protocol would use blocks of 16x16 pixels, corresponding to the native jpeg processing size. Alice would choose a RAW block either from RAW datas $K$ or $C$, to encode a ``0'' or a ``1'' respectively. To decode the message, Bob would compares the image to his key image (also jpeg) block by block.}
\label{fig:Giulia_RGB}
\end{center}
\end{figure*}

\begin{figure}[htbp]
\begin{center}
\includegraphics[width=\columnwidth]{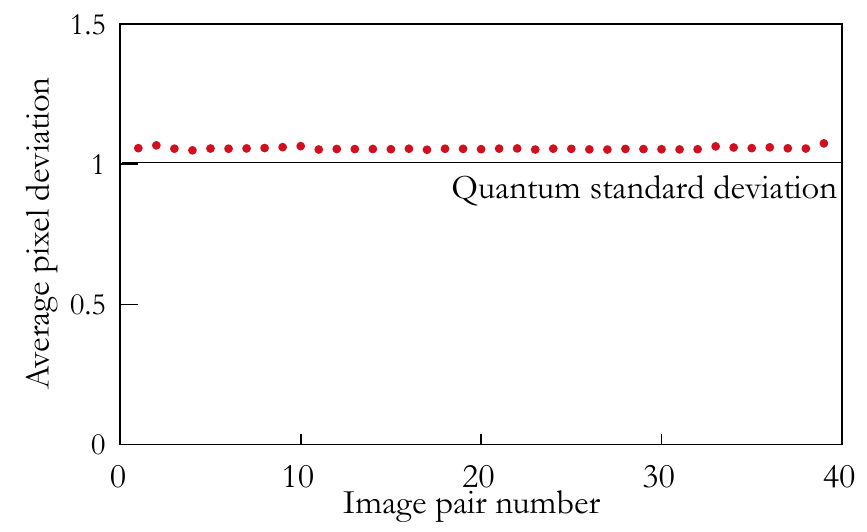}
\caption{Representation of the mean difference between corresponding pixels of two consecutive images, normalised to the expected quantum deviation, which is the square root of the photon number absorbed by the pixel. The mean normalised deviation is $1.05\pm0.1$. }
\label{fig:pixel_deviation}
\end{center}
\end{figure}

\subsection{Protocol Implementation in Color and JPEG}

We have also partially implemented a protocol similar to the one described above. Photographs taken by a commercial colour camera are employed and these images are published in a compressed file format, such as JPG. This significantly increases he complexity of the system and limits the amount of data that can be embedded.
\begin{figure}[htb]
\begin{center}
\includegraphics[width=3cm]{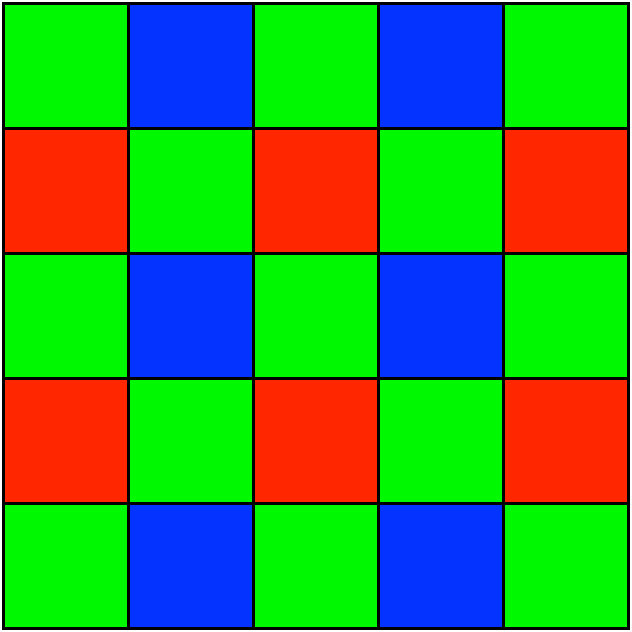}
\caption{Bayer pattern, used in most colour image sensors. Each pixel only measures a single color channel, its full RGB value is extrapolated from adjacent pixels (``debayering''), however the value of it's own color channel is not modified by the algorithm, e.g. the $R$ value of a red pixel is not affected by debarring.}
\label{fig:Bayer_pattern}
\end{center}
\end{figure}
Most colour image sensors are arranged in a \emph{Bayer pattern}, as shown in Fig~\ref{fig:Bayer_pattern}. It consists of the red (R), the green  (G), and the blue (B) components, which are not captured at the same site, but at different places. Every pixel, however, requires each of a R,G and B component to display the appropriate colour. This is done through an interpolation process. For example a red pixel retains its R value, but has associated the G and the B values, resulting from the interpolation of adjacent pixels. If any operation is applied \emph{after} such interpolation process, a.g. white balance, this will introduce some correlations across the pixels and across the colour channels.

For the steganographic technique presented here, the data embedding process (Step 5. above) must only act on the raw image data. All the other processes must happen at a later stage.

When no compression is applied, image processing steps such as white balance and colour correction are reversible, within numerical precision. Step 6. requires an estimator which from a processed image will estimate the most likely values for the captured sensor data for each R, G and B pixel. The comparison between the key-image $K$ and the estimated pixel values from the published stego-image $S$ cannot be a strict equality, but has to use an optimal ``similarity'' bound.
Even with this bound, we found that stronger error correction has to be applied, typically 5\%.

The JPEG image compression algorithm acts on 16x16 pixel blocks. When such compression is used, it is possible to embed a single pixel or several pixels per block. If that is the case, then in Step 6. the block are compared, rather than the pixels. This results in an information capacity of $1/256$ or less, as shown in Fig.~\ref{fig:Giulia_RGB}.

\section{Open Problems}
\emph{``Beware of bugs in the above code; I have only proved it correct, not tried it.'' Don Knuth}

Further research is needed to define a threshold above which assumptions (\emph{i}) and (\emph{ii}) can be considered satisfied.
Below, we present an assessment on the difficulty of this task.

Assumption (\emph{ii}) is easily satisfied: we have measured the correlation in the noise between adjacent pixels and found out that it is unmeasurably low. However, if this remained a worry, it would be possible to use a subset of non-adjacent pixels, leading to a even lower risk of revealing some correlation. Fig.~\ref{fig:autocorrelation} shows that there is no measurable correlation between adjacent pixels.

On the other hand, Assumption (\emph{i}) is demanding: the camera and scene are in general not in the same state when two consecutive photographs are taken. For example, a flying bird in the background or varying illumination would make the stego-image identifiable as such. Furthermore, the image noise would have to be of quantum origin. This means that non-illuminated pixels, which present only classical noise, would have to remain unused during the protocol. In fact, their employment would lead to a major risk of exposing the communication.
In principle, however, if larger areas of a photographs are taken to encode a bit 0 or 1, the protocol does become secure. In the limit that each photograph encodes a single bit, the protocol is obviously secure, and several bits can be sent by using several photographs. Further work would be required to find at what the right compromise would be.
For 40 image pairs, we measured the average difference in the corresponding pixels between the two photographs, and compare it to the expected deviation given only quantum noise. That is, we plot the mean of $(K_i-C_i)/\sqrt{K_i}$, where $K_i$ and $C_i$ are similar, and normalised to represent photon numbers, so $\sqrt{K_i}$ is the expected quantum noise. Results are shown in fig.~\ref{fig:pixel_deviation}.

These differences arise from several factors, one of which is varying image illumination: we have tested a Canon Single Lens Reflex camera, an ATIK scientific camera and a Nokia mobile telephone camera, under several lighting conditions. Taking two consecutive pictures of the same subject, the exposure changes between 0.01\% and 5\%. 
Small differences can be given by effects which are as subtle as sound pressure level, as shown in~\cite{Davis2014}.

\end{document}